# A Survey on Forced Oscillations in Power System

Mohammadreza Ghorbaniparvar

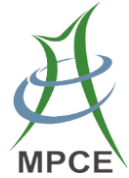

**Abstract** Oscillations in a power system can be categorized into free oscillations and forced oscillations. Many algorithms have been developed to estimate the modes of free oscillations in a power system. Recently, forced oscillations caught many attentions. Techniques are proposed to detect forced oscillations and locate their sources. In addition, forced oscillations may have negative impact on the estimation of mode and mode-shape if they are not properly accounted for. To improve the power system reliability and dynamic properties, it is important to first distinguish forced oscillations from free oscillations and then locate the sources of forced oscillations in timely manner. The negative impact of forced oscillation can be mitigated when they are detected and located. This paper provides an overview on the analysis technique of forced oscillations in power systems. In addition, some future opportunities are discussed on forced oscillation studies.

## 1 Introduction

One of the major threats to the security and stability of power systems is low-frequency oscillations [1], [2]. If oscillations are negatively damped, the magnitude of oscillations will grow and consequently they can cause system breakup and power outages. This is known as small signal stability problems. Analysis of low-frequency oscillations is critical to reliable operation of power systems. Based on the cause of oscillations, low-frequency electromechanical oscillations can be categorized into two main categories: 1) free oscillations and 2) forced oscillations. Free oscillations result from the natural interaction among dynamic devices. In contrast, forced oscillations refer to system responses to an external periodic perturbation.

Methods to mitigate the free oscillations are totally different than those for forced oscillations. Power system stabilizer (PSS) is a simple and economical method to suppress the oscillations by increasing the damping ratio. Control strategies can be utilized to rectify lightly-damped or undamped free oscillations. In contrast, removing the source of oscillations can be considered as a remedy for forced oscillations.

It is very important to get the accurate information about oscillations because it can help operate a power system reliably at its full capacity. To study free oscillations, modal properties of power system provide principal information for control strategies. Basically, there are two approaches for estimating power system modes: 1) component-based approach and 2) measurement-based approach. Stable modes of the power system can guarantee that free oscillations will dampen. On the other hand, undamped modes can be hazardous for the power system. In the past decades, remarkable efforts has been devoted to estimate the modal properties of a power system using component-based approach and measurement-based approach [3]. Numerous measurement-based modal analysis algorithms has been proposed. Valuable overview of existing algorithms are presented in [4], [5].

With the application of phasor measurement unit (PMU) data, it is possible to record and monitor systems' dynamic behaviors. PMU data have been widely used to study the forced and free oscillations because they have a typical sampling rate of 30-60 samples/second and are well synchronized with Global Positioning System (GPS) clock.

One of the primary identification of forced oscillations goes back to 1966 [6]. Forced oscillation can be caused by an external periodic disturbance or a mistuned generator controller. Forced oscillations are reported as a sinusoidal signal originated at generator sites [7]. The resonance between forced oscillation and electromechanical mode can cause system break-down [8]. Sustained oscillation is imposed in power system when forced oscillations exist around a well-damped mode. Forced oscillations have been observed several times in the western North American Power system (wNAPS) with frequency range of 0.2 Hz to 2 Hz which, at some cases, caused moderately hazardous situation [9]. To analyze the forced oscillations, one should initially study the source of forced oscillations in isolated and large systems. Identifying the possible source of forced oscillations can help to attain a more reliable power system. To improve the power systems reliability and stability, detecting and locating the forced oscillations are important. In addition, amplitude, frequency and phase of forced oscillations are of interest.

In this paper, we presented an organized treatment for forced oscillations. Mainly, our objective is three-fold and they are listed below:
1) To provide a comprehensive overview of existing method to detect and locate the forced oscillations and study the impacts of forced oscillations in power systems.
2) To identify the key open problems for detecting the forced oscillations.
3) To introduce future opportunities that can be the possible paths to study the forced oscillations.

Therefore, a clear picture of strength and challenges of the existing methods to study forced oscillations is provided.

The paper is organized as follows. In Section 2, sources of forced oscillations are discussed. Section 3 provides an overview of existing methods to detect the forced oscillations. Oscillations classification methods are discussed in Section 4. Section 5 provides an overview of existing methods to locate the forced oscillations. In Section 6, the impacts of forced oscillations on estimating electromechanical modes are reviewed. Future opportunities for studying forced oscillations are discussed in Section 7. Conclusions are drawn in Section 8.

## 2 Source of forced oscillations

It is important to locate the sources of forced oscillations because once their sources are identified, the forced oscillations can be mitigated by removing the sources. However, it is very challenging to locate the real disturbance source due to the complexity of a power system.

Forced oscillations have been observed and reported in power grids. One of the earliest studies on forced oscillations dates back to 1966 when Ness [6] studied the responses of power systems to cyclic load variations. To study the situation where frequency of forced oscillations is close to natural frequency of generator rotor oscillations, Vournas et al. [8] introduced cyclic load, low-speed diesel generators and wind turbines, which is driving the synchronous generator, as possible sources of forced oscillations. The Pinneilo et al. [10] cited the nuclear accelerator as source of forced oscillations. Power systems behaviors for steady state analysis, dynamic simulation and controller design have been considered in [11], where steel plants, nuclear accelerators, cement mills and aluminum processing plants are cited as sources of forced oscillations. Putting limit on generator field voltage in a single-machine infinite-bus system can induce stable limit cycle that consequently is a cause of forced oscillations via Hopf bifurcation [12]. To analyze the characteristics of regulating systems, electro-hydraulic regulating system of steam turbine is mentioned as source of forced oscillations in [13]. Poorly designed PSS can induce amplification of forced oscillations [14]. Authors of [14] noted that traditional PSS control loop introduces additional poles and zeros which can resonate with cyclic inputs. To avoid possible resonance, authors in [14] provided adjusted PSS design. Rostamkolai et al. [15] proposed that the impacts of cyclic load which is caused by a synchrotron in the power systems is significant, especially in resonance situation with electromechanical oscillations. In [16], incomplete islanding has been cited as possible source of forced oscillations in the power system. To simulate the forced oscillations, mechanical power of generator is modulated with a sinusoidal signal in [17]. These studies showed that forced oscillations may come from different sources.

Reported forced oscillations in these studies are sorted by their reported years and summarized in Table. 1. It can be observed from Table. 1 that:

1) Before year 2000, there are 7 papers studying causes of forced oscillations, i.e., (1)-(7):
- 6 of them used simulation data and they are focused on the load side, i.e., (1)-(5), (7).

2) After year 2000, 20 events have been reported, i.e., (8)-(27).
- 18 oscillation are recorded using PMU data, i.e., (1), (2), (12)-(27).
- 14 oscillation studies are related to generators, i.e., (8), (9), (12)-(21), (23), (27).
- 4 oscillation events are associated with wind power plants, i.e., (12), (17), (19), (21).
- 4 oscillation studies are associated with hydro power plants, i.e., (11), (15), (20), (22).

3) Of all the events, the control actions of 5 events involve the reduction of asset usage (6), (9), (10), (12), (21).

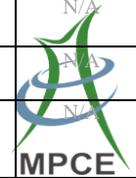

**Table 1.** Forced oscillations sorted by their reported years.

| ID. | Ref Index | year | Reported source | Osc Freq | Control action | Osc Duration | Osc Location | Osc Mag | Osc Signal |
|---|---|---|---|---|---|---|---|---|---|
| 1 | [6] | 1966 | Cyclic load | N/A | N/A | N/A | Simulation studies | N/A | N/A |
| 2 | [10] | 1971 | Nuclear accelerator | N/A | N/A | N/A | Simulation studies | N/A | N/A |
| 3 | [11] | 1988 | Steal plant, nuclear accelerator, cement mills and aluminum processing | N/A | N/A | N/A | Simulation studies | N/A | N/A |
| 4 | [14] | 1990 | PSS on a generator | N/A | N/A | N/A | Simulation studies | N/A | N/A |
| 5 | [8] | 1991 | Cyclic load, low speed diesel generators and wind turbine | N/A | N/A | N/A | Simulation Studies | N/A | N/A |
| 6 | [18] | 1992 June | Failure of an insulator, line switching and system re-configuration (Hopf bifurcation) | 1.0 Hz | Reduce the MW of a generation => stops oscillation | 37 minutes | Rush Island, (East) | 280 MW | Real power |
| 7 | [15] | 1994 | Synchrotron | 0.3-7.5 Hz | N/A | N/A | Simulation studies | 20 MW | N/A |
| 8 | [12] | 2005 | Limits on PSS and AVR of a generator | N/A | N/A | N/A | Simulation Studies | N/A | N/A |
| 9 | [19] [20] [21] | 2005 Nov | Steam Extractor Valve At a Co-Gen plant at Nova Joffre at Alberta | 0.28 Hz | Reduced the steam supply to the industrial process | 6 minutes | Alberta BPA CAISO (West) | 20 MW (Source) 200 MW (COI) | Real Power |
| 10 | [21] | 2008 Jan | Pacific DC Intertie oscillations | 4.0 Hz | Ramp down the PDCI power flow to zero | 55 minutes | BPA, SCE, CAISO (West) | 150MW 200 MVA | Real and Reactive power |
| 11 | [13] | 2010 | Electro-hydraulic regulating system of steam turbine (PSS) of a generator | 0.2-2.5 Hz | N/A | N/A | Simulation Study | N/A | N/A |
| 12 | [21] [22] | 2011 April | Two types of wind turbine during high wind output, Wind power plant | 13.33 Hz | Curtail wind during high wind | 25 minute | OG&E Oklahoma (Eastern) | 5.1 kV | Voltage Mag |
| 13 | [21] | 2011 | Lower a voltage setting at a nuclear generator (voltage setting issue) | N/A | Increase the voltage control back to normal setup. | 12 minutes | Dominion (Eastern) | 250 MW | Real Power Voltage Mag |
| 14 | [21] | 2012 before | Malfunction of the governor controller at a large coal-fired generating plant during tests | 0.285 Hz | N/A | 2.5 minute | MISO (Eastern) | 40 MW | Real Power |
| 15 | [23] | 2013 summer | Hydro Generator (Vortex control, Rough zone during startup/shut down) | 0.38 Hz | Move the Hydro Generation out of its rough zone | Several hours | WECC | 10 MVA | Apparent Power |
| 16 | [24] | 2013 April | A Generator | 0.12 Hz | N/A | N/A | NE ISO (Eastern) | 100MW | Real Power |
| 17 | [25] | 2013 April | Wind plant | 13 Hz | N/A | Several hours | BPA (West) | N/A | Real Power |
| 18 | [26] [21] | 2013 May | AVR Controller malfunctions of a generator | 1.25 Hz | Turn off excitation system | 2 minutes | NYISO (Eastern) | N/A | Voltage Magnitude, Power |
| 19 | [21] | 2014 before | A controller of a wind plant | 14 Hz | The manufacturer upgraded the plant control | N/A | BPA (Eastern) | N/A | Real and Reactive Power |
| 20 | [21] | 2014 Feb | A faulty control card for a hydro power plant | 1.80 Hz | Replaced the faulty control card | N/A | ERCOT | N/A | Frequency |
| 21 | [21] | 2014 | Controller setting of a wind plant | 3.3 Hz | Reduce wind power plant outputs (later, restore controller settings) | 17 hours | ERCOT | N/A | NA |
| 22 | [24] | 2014 | N/A | 0.14 | N/A | N/A | NE ISO | 20MW | Real Power |

| | | Feb | | Hz | | | (Eastern) | | |
|---|---|---|---|---|---|---|---|---|---|
| 23 | [27] | 2014 Oct | Surging water vortex on turbine at a hydro generator | 0.33 Hz | N/A | >1 minutes | BPA (Western) | N/A | Real, Reactive Power |
| 24 | [21] | 2015 before | Load induced oscillation (A Refinery) | 4.62-5.0 Hz | N/A (Use IEEE 519 standard) | 15 minutes | OG&E (Oklahoma) (Eastern) | N/A | Voltage Mag |
| 25 | [24] | 2015 Mar | N/A | 1.48 Hz | N/A | N/A | BPA (West) | N/A | |
| 26 | [24] | 2015 Nov | N/A | 1.17 Hz | N/A | N/A | BPA (West) | N/A | |
| 27 | [27] | 2015 Oct | Interaction between an under-excited limiter and a PSS of a generator (Dalles) | >1 Hz | Move the units out of the under excitation area | >60 minute | Dalles BPA (West) | N/A | Real, Reactive Power |

4) 13 oscillations have the records on their time durations:
- 4 oscillations last for more than 1 hours, i.e., (15), (17), (21), (27).
- 5 oscillations last for more than 10 minutes but less than 1 hours, i.e., (6), (10), (12), (13), (24).
- 4 oscillations last for less than 10 minutes, i.e., (9), (14), (18), (23).

5) 9 oscillations have the records on their oscillation amplitudes in MW or MVA:
- The smallest peak-peak oscillation magnitude is 10 MVA, i.e., (15).
- The largest peak-peak oscillation magnitude is 280 MW, i.e., (6).
- 5 oscillations have their magnitude larger than 100 MW, i.e., (6), (9), (10), (13), (16).
- 4 oscillations have their magnitude less than 40 MW, i.e., (7), (14), (15), (22).

6) 19 oscillations have the records on their oscillation frequency in Hz. According to [46], they are categorized as follows:
- 2 oscillation between (0.01-0.15Hz), i.e., (16), (22).
- 7 oscillations between (0.15-1.0 Hz), i.e., (7), (9), (11), (14), (15), (16), (23).
- 7 oscillations between, i.e., (1.0–5.0 Hz), i.e., (6), (10), (18), (20), (21), (25), (26).
- 4 oscillations between (5.0-14.0 Hz), i.e., (12), (17), (19), (24)

7) 19 oscillations have the records on their location sources:
- 9 oscillations are recorded in Eastern interconnection, i.e., (6), (12)-(14), (16), (18), (19), (22), (24).
- 8 oscillations are recorded in Western interconnection data, i.e., (9), (10), (15), (17), (23) (25)-(27)

It can be observed that both cyclic load and generator control equipments (GCE) were reported as the sources of forced oscillations. In addition, in the recent years, GCEs such as PSS and automatic voltage regulators (AVR) are reported more often as the sources of forced oscillations. In other words, the sources of the forced oscillations mainly are located on the generator sides. This prior information suggests that to detect and locate forced oscillations, one may want to place more attentions to the generation sides.

## 3 Oscillation detection methods

To provide the better understanding of forced oscillations, Fig. 1 illustrates sample time plot of forced oscillations in simulated PMU data. Assume that the forced oscillations are modeled by a sinusoidal signal with unknown amplitude, frequency, phase and arrival time embedded in colored noise. In (1), A is a sinusoidal signal which represents forced oscillations. Symbol $z_t$ is the additive noise and N is the total number of samples.

$$x_t = A + z_t \qquad t = 1,2,...,N \qquad (1)$$

Therefore, the challenge is to detect the sinusoidal covered with ambient noise.

In [28], an algorithm to detect the sinusoidal signal under Gaussian white noise (GWN) is presented. Forced oscillations can be detected using statistical properties of periodogram. To detect sinusoid covered with noise, simple estimators of power spectral density (PSD) can be used. PSD describes how the strength of a signal is distributed over frequency. Using the Welch method, Fig. 2, Fig. 3 and Fig. 4 illustrate PSDs of $x_t$ for the 30 seconds, 1 hour and 24 hours of data consequently with window length = 128, Hamming windows, and overlapping = 50%. As it can be seen in Fig. 2, there are four peaks. Peaks at 0.4 HZ, 3 Hz and 9 Hz represent the modes from the ambient noise. Peak at 5 Hz shows forced oscillations. This peak can be smaller for forced oscillations with lower SNR. However, Fig. 3 illustrates just two peaks, one at 0.4 Hz (i.e., mode from the ambient noise) and another at 5 Hz with high amplitude (i.e., the forced oscillations). Fig. 4 clearly shows only one peak at 5 Hz with a very high amplitude. This observation indicates that there is a direct relation between the magnitude of forced oscillation and time, which can be used to distinguish the different types of oscillations. Using PSD to detect the forced oscillation, there is a trade-off between time and the detection accuracy. Because it is important to detect the forced oscillations in a timely manner, using PSD to detect the oscillations is not the best practice. Therefore, there is a need for a method that can distinguish different types of oscillations in a timely manner.

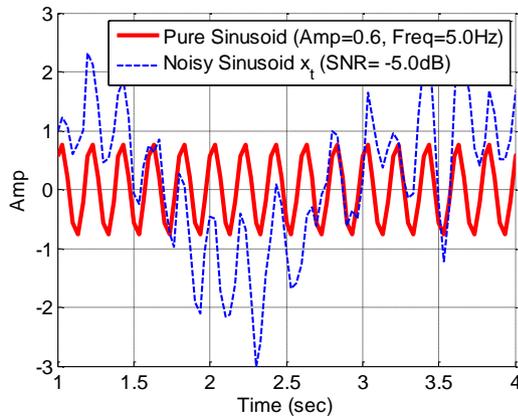

**Fig. 1** Time plots of existing forced oscillations in PMU data and its pure sinusoidal components.

Because ambient data in power systems are known to be colored, it is not practical to directly apply the proposed method in [28]. Thus, Follum et al. [7] modified the method to deal with the colored ambient noise environment. They proposed methods for estimating the frequency, amplitude, start time and end time of forced oscillations. Yet, the problem of detecting oscillation in a timely manner, as mentioned for PSD, remains unsolved. It is expected to observe a close correlation between the location and the source of forced oscillation (in a large system) with the detection accuracy. Proposed method in [7] is only applied to the simulated data rather than simulation data from power system or field measurement data. As a result, there is a question on real-word application of the proposed method. To decrease the time-delay for detecting forced oscillations, the coherence spectrum, also known as magnitude-squared coherence, is employed to detect forced oscillations in [29]. It reflects the linear correlation of time-series signals at a specific frequency. Although it decreased the time-delay for detecting forced oscillations even for small signal-to-noise ratio (SNR), it has two drawbacks: 1) it requires two channel of data and 2) it may introduce some peaks because of linear correlation between two channels. The method is further developed in [17] as a self-coherence method. Fundamentally, the self-coherence spectrum is the coherence spectrum using one channel data and its time-delayed signal. The self-coherence method is applied to the data set with a time delay of $\Delta t = 6.0$ s. For the 60 minutes of date, Fig. 5 shows the self-coherence spectrum in a heat map where oscillations frequency is 8 Hz.

During the ring-down oscillations, using self-coherence method, existence of time delay eliminates the effect of the linear correlation, which causes the peaks in the coherence method. Yet, the problem of proper channel selection in [17] remains unsolved. In addition, this method does not specifically detect forced oscillations, rather it provides a magnitude of coherence between 0 to 1. Furthermore, a bootstrap method was applied to set up appropriate threshold on self-coherence for detecting forced oscillations in [30]. Based on the bootstrap method, a threshold can be determined to detect the sustained oscillations with pre-selected probability of false alarm. Fig. 6 shows the threshold as a red line as with a false alarm rate set at $\alpha = 1\%$. As it can be seen the only point that exceeds the threshold is at frequency of 8 Hz. However, there is question of how quickly this method can detect oscillations. To extend the coherence method, Zhou [31] applied coherence method to multiple-channel data. It shows that using multiple-channel method can increase detection rates. To inspect the performance of coherence method, different algorithms such as Welch algorithm, Capon algorithm and autoregressive moving average (ARMA) algorithm are used to estimate coherence spectra. Moreover, performance of three algorithms are compared. However, there is a question on applicability of the method when both forced and free oscillations exist in the system. In [23], covariance-driven stochastic subspace identification (SSI-COV) [32] was applied to detect the forced oscillations. This method can simultaneously detect forced oscillations while estimating system modes. The authors studied the effects of forced oscillations on the local modes and inter-area modes. However, proposed method fails to distinguish forced oscillations from free oscillations.

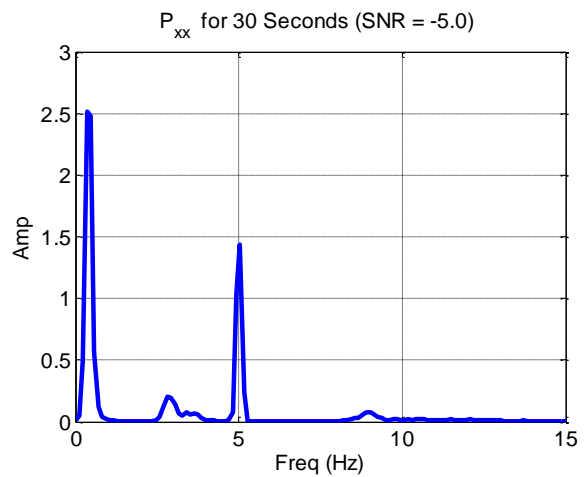

**Fig. 2** The PSDs of $x_t$ for the 30 seconds of data in the simulation data.

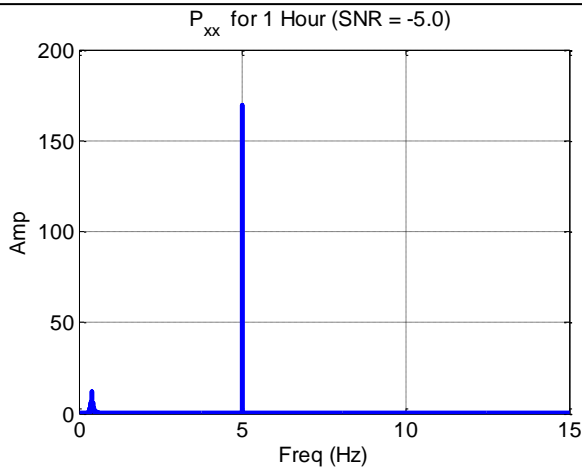

**Fig. 3** The PSDs of $x_t$ for the 1 hour of data in the simulation data.

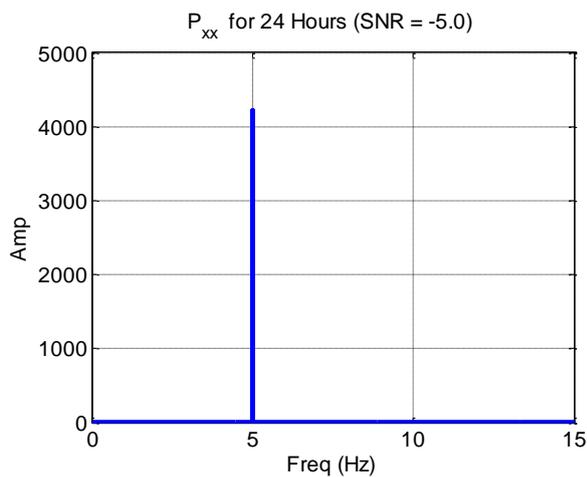

**Fig. 4** The PSDs of $x_t$ for the 24 hours of data in the simulation data.

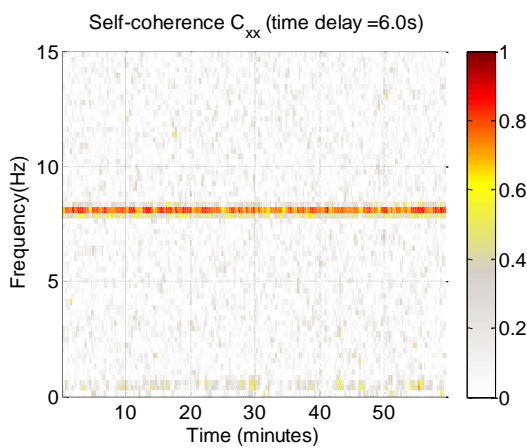

**Fig. 5** Self-coherence spectrum (with a peak at oscillation frequency of 8 Hz) for 60 minutes of the simulation data.

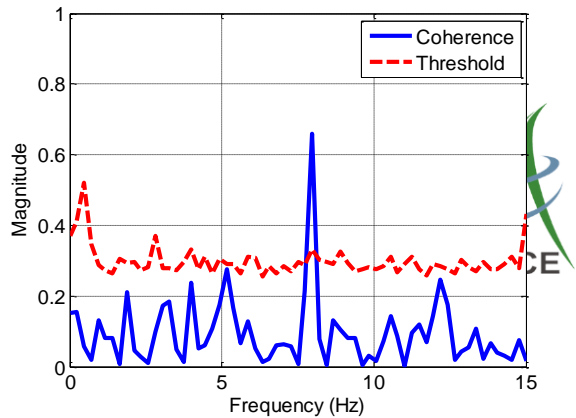

**Fig. 6** Self-coherence spectrum $C_{xx}$ ($\Delta t = 6.0$s) and estimated threshold ($\alpha$=1%) for the first 34+6 s in the simulation data.

## 4 Oscillation classification methods

Oscillations classification is very important to the reliable operation of power systems because control strategy for suppressing different types of oscillations can be significantly different. Using support vector machine (SVM), Liu et al. in [33] proposed a method that can classify two different types of oscillations based on feature extraction. Basically, increasing amplitude of free oscillations is used to distinguish it from forced oscillations. In this case, algorithm accuracy has direct relationship with envelope size. Thus, there is a question of whether this algorithm can work in real-time scenarios. Xie et al. in [34] proposed unique mathematical models for forced and free oscillations. In addition, they showed that the responses of the system to forced oscillations differs from those of free oscillations. Based on noise response and harmonic nature of oscillations, authors proposed a method to distinguish forced oscillations from free oscillations. However, there is question of how to separate the pure sinusoid form noise in both cases (free and forced oscillations) and the effects of this separation on algorithm accuracy should be considered. In addition, it is important to evaluate the performance of proposed method when forced oscillations exist at system's mode frequency.

To study the fundamental methodologies adopted for oscillation classifications, the two-area model shown in Fig. 7 is utilized to generate the data for both forced and natural oscillations. Note that Power System Toolbox (PST) [35] is used to simulate the data with a classical generator model. This model has three modes that are presented in Table. 2. Active and reactive loads of all the buses is modulated by 5% GWN to simulate the ambient noise. To generate forced oscillations (Fig. 7.a), a sinusoidal signal with amplitude of 0.1 is injected into the system by modulating the shaft torque of *G1* for 5 minutes. To generate natural oscillation (Fig. 7.b), generation values are set up to reach an unstable condition after a three phase fault occurs between the bus 7 and 8.

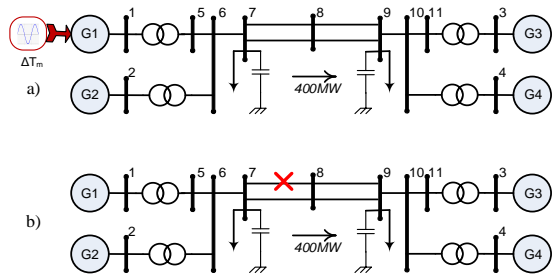

**Fig. 7** The two-are model with a) forced oscillation and b) natural oscillations.

**Table 2.** The modes of the two-area model.

| Mode Index | Frequency(Hz) | Damping Ratio (%) |
|---|---|---|
| 1 | 0.53 | 7.6 % |
| 2 | 1.14 | 3.4% |
| 3 | 1.18 | 3.5% |

We conducted a study for four different cases: 1) Natural oscillations exist at Frequency of 0.4Hz; 2) Forced oscillations exist at frequency of 2Hz; 3) Forced oscillations exist at frequency of 1.14Hz (close to system's mode frequency); and 4) Forced oscillations exist at frequency of 0.53Hz (close to system's mode frequency). The spectral approach, proposed method in [34], is applied to these cases and results are shown in Table. 3. Here, $\alpha_S$ indicates the ratio of PSD between two different channels when oscillations exist where $\alpha_N$ indicates the ratio of PSD of two different channels when oscillations does not exist

**Table 3.** Simulation results using[34].

| Case Index | 1 | 2 | 3 | 4 |
|---|---|---|---|---|
| *Oscillations Frequency (Hz)* | *0.4* | *2* | *1.14* | *0.53* |
| $\alpha_S$ | 0.0017 | 2122.9 | 105.38 | 0.19 |
| $\alpha_N$ | 0.2544 | 3.18 | 132.89 | 0.36 |
| *Actual Oscillation type* | Natural | Forced | Forced | Forced |
| *Spectral Approach*[34] | Natural | Forced | Natural | Natural |

As it can be seen in Table. 3, $\alpha_S$ and $\alpha_N$ are very close for case 1 so it can be considered as natural oscillations. In addition, the PSD ratios at 2 Hz are NOT equal which indicates that the forced oscillations exist. However, if forced oscillations' frequency is close to mode frequency of the system (i.e. 1.14 and 0.53 Hz) as in case 3 and 4, the PSD ratios are close to each other. In other words, the spectral approach [34] is not effective when forced oscillations exist at system's mode frequency.

Wang et al. [36] introduced three different mathematical models for weakly damped oscillations, limit cycle and forced oscillation. Using Kurtosis estimation and PSD estimation, they proposed an algorithm which can distinguish these oscillations from each other. Authors of [36] used standard deviation of PSD as a metric to detect forced oscillations. However, this method just has off-line application. To get the accurate PSD estimation for distinguishing oscillations, as Fig. 2, 3 and 4 illustrated, reasonable amount of time required. In addition, this method may not be effective for oscillations with small signal-to-noise ratio (SNR).

As it can be seen, forced oscillations play a critical role in system performance and they can cause catastrophic events [1], [2]. Therefore, to increase the reliability of power systems, it is important to detect and distinguish forced oscillations and free oscillations. A robust method which can distinguish forced oscillations from free oscillations even when forced oscillations exist at system's mode frequency is needed. Extracting the unique features of forced oscillations can be a possible solution. However, unknown nature of forced oscillations source has made it hard to model them mathematically.

## 5 Oscillation source locating methods

In order to suppress the forced oscillations and mitigate their impacts, a robust method for locating the sources of forced oscillations is needed. Fundamentally, forced oscillations can be caused by an external perturbation in generator sites.

PMU data reflects the behaviors of the model at 30 to 60 samples/s. They basically are phasor measurements synchronized with GPS clock. PMU data have been widely used in power systems analysis [37],[5]. In [38], the authors proposed new application of PMU data. PMU data have been utilized to locate the sites of disturbance sources. A hybrid dynamic simulation method is used to locate the forced oscillations. Basically, the idea is to replace part of dynamic states and algebraic variables with measurements, which in this case is PMU data, and run power system model. Mismatch between simulation data and PMU data indicates the location of forced oscillations. Although the method shows promising results, it is only applicable in post-fault analysis studies. Using dynamic state variables, an energy based method is applied to locate the source of forced oscillations in [39], [40]. Although both methods shows promising results, it requires detailed information of the model which makes the applicability and the validity of this method unclear. In [41], after visually detecting oscillations, Prony analysis is applied to extract the oscillations' characteristic (i.e. frequency, Magnitude, Phase and Damping ratio) and consequently damping torque of the generator is used to locate the source of forced oscillations. The authors detected the forced oscillations at frequency of 0.56Hz which is close to the mode frequency and the source of oscillation is not discussed in the paper.

Thus, it is interesting to study the performance of this method for two different scenarios: 1) Forced oscillations with very small frequency (i.e. 0.13) imposed on a generator and 2) applicability of method on a large power system. Li et al. in [42] analyzed the forced oscillations at 0.7475Hz based on extended equal-area criterion (EEAC) theory and then locate the forced oscillations in two steps: 1) identify the disturbance cluster by comparing steady phase relationship, 2) locate the disturbance source by comparing transient response trajectories phase of observation points in identified disturbance cluster. However, the applicability of proposed method for forced oscillations with high frequency (i.e. 14Hz) is questionable. In [43], Yu et al. studied the differences between the free and forced oscillations in terms of their energy conversions. Considering the cyclical load disturbance as a cause of forced oscillations, they proposed that an increase of potential power can be used to locate the disturbance source of forced oscillations. As it is listed in section 2, the cyclical load disturbance is one of the possible source of forced oscillations besides malfunctioning generator, PSS limit on generators, nuclear accelerator and etc. So, it is important to evaluate the performance of proposed method in [43] for different sources (especially when forced oscillations' frequency is considerably small). Hu et al. [44] proposed a new approach based on power dissipation to locate the disturbance source of forced oscillations. As it can be seen in Table. 1 from reported event, the forced oscillations' frequency can be expected to be between 0.12Hz to 14Hz and it is not necessarily equals to natural oscillations' mode frequency. Thus, the performance of the proposed method is questionable when forced oscillation's frequency is not equal to system's mode frequency. In [23], it was proposed that the location of forced oscillations can be determined using mode shape magnitude when there is not a strong resonant effect. However, in the case of resonance when frequency of forced oscillations is close to that of system modes, mode shape estimation may not be useful to locate the source of forced oscillations.

To provide the better understanding of forced oscillations locating methods, power spectrum density of the voltage of each buses for case 2 and 4 in Section 4 is calculated and plotted in Figs. 8 and 9. It can be observed in Fig. 8 that the PSD of magnitude response of buses are consistent with their distance to the oscillation source. In other word, more we get closer to oscillation source more the magnitude of PSD will be in oscillation frequency. Here, we see bus number 1 and 5 has highest magnitude and bus number 3 and 4 has lowest magnitude which indicates oscillation is located in generator 1. However, this method can locate the forced oscillations with small frequency. As it can be seen in Fig. 9, the PSD of magnitude response of buses are not consistent with their distance to the oscillation source when forced oscillations has small frequency (i.e.

0.53 Hz). Thus, it is important to find a method that can locate the source of forced oscillation at any frequency even close to system' mode frequency.

While aforementioned methods were able to locate the forced oscillations to some degree, there is a need to develop a method which can distinguish forced oscillation from free oscillations and locate the source of the forced oscillations under different scenarios: 1) forced oscillations with different frequency range (i.e. 0.12Hz to 14 Hz), 2) forced oscillation in a large and complex system, 3) forced oscillations with low SNRs and 4) forced oscillations imposed with different causes (i.e. Cyclical load, generator malfunctioning and etc).

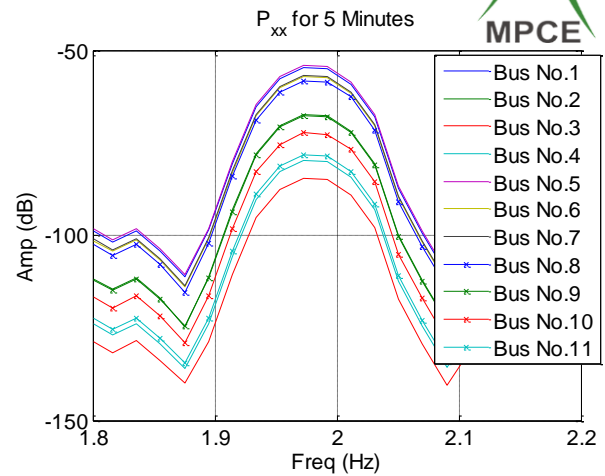

**Fig. 8** The PSDs of $x_t$ for 5 minutes of data for all the buses for case 2 Section 4.

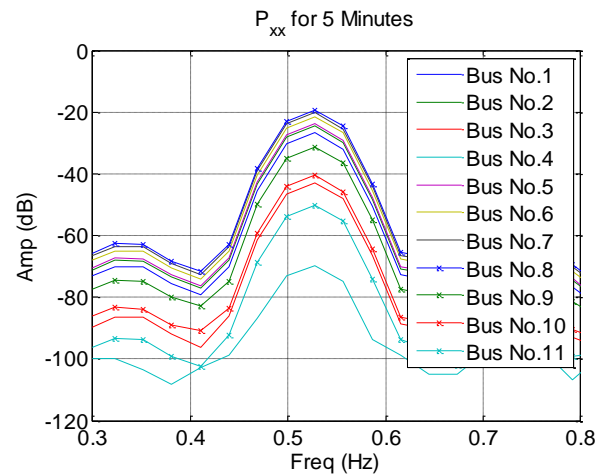

**Fig. 9** The PSDs of $x_t$ for 5 minutes of data for all the buses for case 4 Section 4.

# 6 Effects of forced oscillation on mode estimation algorithms

Modes and mode shapes are critical parameters describing the free oscillations. Mode describes the system's oscillation frequency and damping behaviors while mode shape states the relative participation of state variables in particular mode [45]. Electromechanical modes in power systems are analyzed through many parametric methods.

Basically there are two approaches for estimating power system modes: 1) simulation model-based approach [46] and 2) measurement-based approach. Regarding model-based approach, the challenge is to obtain an accurate dynamic model for real-time power system operating conditions. The task is even more difficult for large-scale power systems. The difference between real system responses and simulation model responses, during a real-world event, indicates the model's deficiency which can cause conservative operation and reducing asset utilization. For instance, an initial effort was made by Kosterev et al. [1] to build a component based model for simulating the Western Electricity Coordinating Council (WECC). The study shows that the simulated system responses do not match the field measurement data. Simulation results and measurement results matched only after extensive modification and calibration. In contrast, the measurement-based approach requires less effort than the model-based approach. In addition, measurement-based approaches can update the mode estimation based on incoming measurement data. Based on measurement approaches, modes of a system can be identified using ambient data, ring-down data and probing signals. Therefore, measurement-based methods have certain advantages over the model-based eigenvalue analysis in monitoring real-time power oscillations. An overview of mode estimation algorithms is given in [4], [5].

To estimate the modal properties of a system, mode-meter algorithms are constantly processing PMU data and estimating frequency, damping ratio and shape of a mode from ambient data. Forced oscillations impose a challenge on mode estimation algorithms [47], [5]. To estimate the mode accurately, it is important to study the impact of forced oscillations on mode estimation. Studying both simulation data and real-time PMU data, authors in [48], pointed out that damping estimate can be comprised when forced oscillations exist. Vanfretti et al. showed that the impact of forced oscillation on damping estimate depends on its location with respect to inter-area mode [49]. It has been shown that estimated damping ratio decreases when a forced oscillation is superimposed on the inter-area mode. However, estimated damping ratio increases when forced oscillations are close to the inter-area mode. Myers et al. showed that existence of forced oscillations can have slight impact on mode-shape estimation only when forced oscillations frequency is close to system mode frequency [9]. Forced oscillations with frequency far away from system mode frequency have no noticeable impact on mode-shape estimation.

Properties of the electromechanical modes describe system oscillation frequency and damping behavior. Therefore, identifying forced oscillations is important because it helps mode estimation algorithms to provide more accurate information about power-system stability.

## 7 Future opportunities for studying forced oscillations

Clearly, detecting and locating oscillations leads to a more reliable power system operation. Fig. 10 gives a summary of papers that studied forced oscillations. As it can be seen, 6 papers studied the detection methods [7], [17], [29]–[31], [23], 8 papers studied the location of forced oscillations [23], [38]–[44], 5 papers studied the effects of forced oscillations on mode estimation algorithms[5][9], [47]–[49] and just 3 papers focused on oscillation classification [33], [34], [36]. Although it is very important to distinguish oscillations, there are not many methods to address this issue.

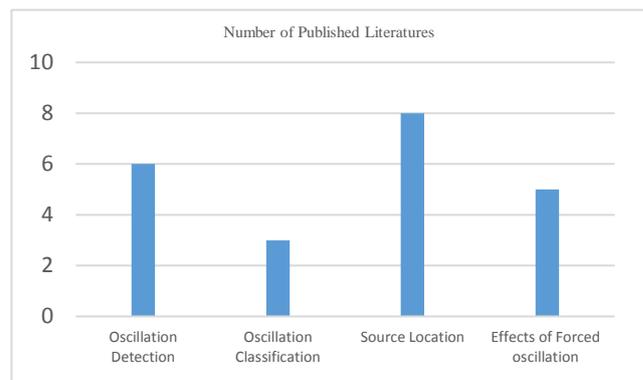

**Fig. 10** Number of published literatures on studying forced oscillations.

Beyond the previously existing methods to locate and detect forced oscillations, many future opportunities exist such as:
- The relation between length of data and magnitude of PSD, as it has been shown in Figs.3, 4 and 5, can be used to distinguish different types of oscillations.
- As it has been discussed throughout the paper, coherence analysis has shown promising results in detecting the forced oscillations. However, there is still some deficiency in the proposed algorithms. Applying other algorithms and improving existing algorithms can be critical for detecting the forced oscillations.
- Effects of data length on PSD estimation to detect the forced oscillations.
- Comprehensive study to reveal the advantage and disadvantage of different algorithms to estimate the coherence spectra for detecting oscillations.
- Distinguishing free and forced oscillations using their distinct features.
- Combination of coherence and self-coherence for a system to monitor the propagation information of oscillation and to locate the sources of forced oscillations.
- PMUs are installed throughout systems. They carry different information based on their location, so their distinct characteristics can be used to detect and locate the oscillations. For example, coherence amplitude and phase difference of PMU data can be used to locate the forced oscillations location.
- One of the important issues in power systems is the placement of PMUs. Because PMU data can be used to locate the disturbance source of forced oscillations, it can be a factor for designing a PMU placement algorithm. For example, some new constraints can be added to the optimization problem of PMU placement to include the considerations of locating forced oscillations.

## 8 Conclusion

This paper presented a comprehensive overview of the forced oscillations in power systems. Basically five major areas are discussed: 1) source of forced oscillations 2) detecting methods of forced oscillations 3) distinguishing methods of forced oscillations 4) locating methods of forced oscillations 5) impact of forced oscillation in power systems. In addition, future opportunities to study the forced oscillations are discussed. It can be concluded that in a large power system, GCEs (such as PSS and AVR) are often reported as the sources of forced oscillations in the recent years. As it has been reported, the resonance between forced oscillations and electromechanical mode can cause catastrophic events. In addition, forced oscillations in a power system have negative impact on mode estimation algorithms. Thus, they should be detected and located in their early stage.

There is a need of a comprehensive algorithm which distinguishes forced oscillations from free oscillations and locate the sources of forced oscillations accurately in a large and complex power system.